\documentclass[%
 reprint,
 amsmath,amssymb,
 aps,
]{revtex4-2}

\usepackage{xspace}
\usepackage{amssymb}
\usepackage{amsmath}
\usepackage[utf8]{inputenc}
\usepackage{graphicx}
\usepackage{cancel}
\usepackage{bbm}
\usepackage{color}
\usepackage{lineno}
\usepackage{dcolumn}   
\usepackage{bm}        
\usepackage{comment}
\usepackage[dvipsnames]{xcolor}
\usepackage[colorlinks=true,
            linkcolor=RoyalBlue,
            urlcolor=RoyalBlue,
            citecolor=RoyalBlue]{hyperref}

\newcommand{\snn}          {\ensuremath{\sqrt{s_{\mathrm{NN}}}}\xspace}

\newcommand{\twoHnn}       {$\sqrt{s_{\mathrm{NN}}}=200$~GeV\xspace}
\newcommand{\twosevensixnn}{$\sqrt{s_{\mathrm{NN}}}=2.76$~TeV\xspace}
\newcommand{\fivenn}       {$\sqrt{s_{\mathrm{NN}}}=5.02$~TeV\xspace}

\let\jpsi=\Jpsi

\begin{document}
\title{Probing Gluon Shadowing in Heavy Nuclei through Bayesian Reweighting of J/$\psi$ Photoproduction in Ultra-Peripheral Collisions}
\date{\today}

\author{Pengzhong Lu}
\author{Zebo Tang}
\email{zbtang@ustc.edu.cn}

\author{Xin Wu}

\author{Wangmei Zha}
\email{first@ustc.edu.cn}
\affiliation{State Key Laboratory of Particle Detection and Electronics, University of Science and Technology of China, Hefei 230026,China}

\begin{abstract}

The gluon distribution in nuclei plays a pivotal role in understanding quantum chromodynamics (QCD) under extreme nuclear environments, yet remains poorly constrained compared to quark distributions. Coherent \jpsi photoproduction in ultra-peripheral heavy-ion collisions ($\gamma + A \rightarrow \mathrm{J}/\psi + A$) provides a unique solution to this challenge, serving as a sensitive probe of nuclear gluon densities. 
In this study, we perform Bayesian reweighting on the EPPS21 and nCTEQ15 sets of nuclear parton distribution functions (nPDF) by incorporating coherent \jpsi photoproduction measurements from both RHIC and LHC. The Bayesian-reweighted gluon modification factors $\mathrm{R_g^{A}}(x, Q^2 = 2.4\ \mathrm{GeV}^2)$ reveal pronounced nuclear shadowing in the Pb nuclei, with $\mathrm{R_g^{\mathrm{Pb}}} \approx 0.60$ at $x = 10^{-4}$, while simultaneously achieving a great reduction of the uncertainties in the density of the gluon across the critical Bjorken-$x$ range $10^{-5} < x < 10^{-3}$ compared to initial predictions of the nPDF.
This work establishes coherent \jpsi photoproduction as a precision tool for gluon nPDF extraction, overcoming traditional deep-inelastic scattering limitations through perturbative QCD-calibrated probes. The constrained nPDFs demonstrate improved consistency with the experimental data across collider energies, particularly in the shadowing-dominated regime.
\end{abstract}

\maketitle

\section{Introduction}

Parton distribution functions (PDFs) quantify the probability of finding a parton carrying a longitudinal momentum fraction $x$ at energy scale $Q^2$. Being non-perturbative objects, PDFs of quarks and gluons cannot be computed \textit{ab initio} but must be extracted through global analyses comparing theoretical predictions with experimental data. Proton PDF determination has evolved into a precision science through systematic studies of hard scattering processes, with modern global analyses~\cite{Dulat:2015mca,Harland-Lang:2014zoa,Accardi:2016qay,Alekhin:2014irh,Alekhin:2017kpj,NNPDF:2017mvq} leveraging advanced perturbative QCD calculations coupled with sophisticated statistical methods.

Extending this framework to nuclei introduces additional complexity: Nuclear PDFs (nPDFs) are modified by collective nucleon-nucleon interactions, manifesting phenomena like nuclear shadowing and gluon saturation. Precise knowledge of nPDFs is indispensable for interpreting high-energy nuclear collision phenomena, particularly the Quark-Gluon Plasma (QGP) studied at RHIC and LHC~\cite{Busza:2018rrf,ALICE:2022wpn,STAR:2005gfr,PHENIX:2004vcz}. Although quark nPDFs can be constrained through lepton-nucleus DIS~\cite{EuropeanMuon:1983wih} and proton-nucleus Drell-Yan~\cite{Stavreva:2010mw} measurements, the gluonic sector remains elusive due to the lack of direct probes sensitive to gluon dynamics.

This challenge can be addressed by coherent \jpsi photoproduction in ultra-peripheral collisions (UPCs), where relativistic heavy ions (Pb, Au) interact exclusively via electromagnetic fields without nuclear overlap. In this process, the photon-Pomeron interaction ($\gamma + \mathbb{P} \rightarrow \mathrm{J}/\psi$) is mediated by two-gluon exchange in QCD language, providing direct access to the gluon structure. The large \jpsi mass ($m_{\mathrm{J}/\psi} \approx 3.1\ \mathrm{GeV}$) enables rigorous perturbative QCD calculations, which establish that the forward production cross-section scales with the square of the gluon density~\cite{Soff:1993qy,Cao:2018hvy}. This quadratic dependence, combined with the coherent enhancement (requiring intact nuclei post-interaction), makes \jpsi photoproduction a uniquely sensitive probe of nuclear gluon distributions.

Recent developments in extracting gluon nPDFs from coherent \jpsi photoproduction have addressed several key challenges. The pioneering work by V. Guzey~\cite{Guzey:2013qza} first proposed utilizing UPC coherent \jpsi production to probe nuclear gluon distributions. By analyzing ALICE UPC measurements at mid-rapidity ($y=0$), this study provided the first evidence of gluon shadowing in Pb nuclei. However, this approach was fundamentally limited to $y=0$ measurements due to the inherent ambiguity in determining the photon source direction in heavy-ion collisions, where both nuclei can serve as photon emitters with identical $\gamma N$ energies at mid-rapidity. To overcome this limitation, J.G. Contreras~\cite{Contreras:2016pkc} proposed a novel method utilizing differential measurements across different impact parameter ranges. By comparing coherent \jpsi yields in peripheral and ultra-peripheral collisions, this approach attempted to disentangle the two photon source contributions. Nevertheless, this method faced challenges from potential nuclear overlap effects in non-ultra-peripheral collisions. A significant breakthrough came from the CMS collaboration~\cite{CMS:2023snh}, which implemented a neutron tagging technique using Zero Degree Calorimeters (ZDCs) to control the impact parameter in UPC events. This innovative approach successfully resolved the two-component ambiguity by separating the photon emission directions, enabling independent extraction of gluon nPDFs from both nuclear sources. The ALICE collaboration subsequently adopted similar methodology.

Despite these advancements, several limitations persist in current analyses. The theoretical modeling of neutron emission probabilities as a function of impact parameter remains uncertain, particularly for higher-order processes:
\begin{itemize}
    \item Simultaneous photon exchange exciting both nuclei
    \item Nuclear excitation accompanying photon emission
\end{itemize}
These effects, currently unaccounted for in most theoretical frameworks, may introduce systematic uncertainties in the extracted cross-sections and consequently affect the precision of gluon nPDF determination.


Given the theoretical challenges in precisely calculating neutron emission probabilities, which appear intractable in the near future, we propose an alternative strategy that directly utilizes existing nPDF parameterizations (EPPS and nCTEQ) combined with Bayesian reweighting techniques to address the two-component ambiguity. This approach circumvents the need for neutron tagging and its associated theoretical uncertainties, providing a complementary method for gluon nPDF extraction while serving as an independent validation of the ZDC-based impact parameter control approach. In this work, we employ the Bayesian reweighting technique~\cite{Giele:1998gw,Watt:2012tq,Ball:2011gg,Sato:2013ika} to refine nPDFs by updating prior distributions with all available \jpsi photoproduction measurements in UPC from RHIC and LHC. Unlike traditional fitting methods, Bayesian reweighting systematically adjusts initial assumptions (priors) to incorporate new observations, leading to more precise constraints on nPDFs. 

The paper is organized as follows. Section~\ref{methodology} details our theoretical framework, including the Impulse Approximation (IA)  framework for coherent \jpsi production in UPCs and implementation of Bayesian reweighting for nPDF updates. Section~\ref{results} presents our comprehensive analysis, where we extract rapidity-differential cross sections for coherent \jpsi production within the IA framework, derive nuclear gluon modification factors, predict \jpsi photoproduction cross sections as functions of $\mathrm{W}_{\gamma \mathrm{N}}$, and compare predictions with RHIC and LHC experimental measurements obtained using neutron tagging techniques. Finally, Section~\ref{summary} summarizes our key findings, and outlines future research directions in this field.

\section{Methodology}
\label{methodology}

\subsection{Impulse Approximation for coherent \jpsi production predictions}

To estimate the cross section for coherent photon-induced \jpsi production in UPCs, we employ a methodology based on the convolution of the Weizsäcker-Williams photon spectrum and the photonuclear cross section~\cite{PhysRevC.60.014903,PhysRevLett.92.142003}. The total cross section for J/$\psi$ production is computed as:
\begin{equation}
\sigma(\text{AA} \rightarrow \text{AA} \text{J}/\psi) = \int d\omega_{\gamma} \frac{dN_{\gamma}(\omega_{\gamma})}{d\omega_{\gamma}} \sigma(\gamma \mathrm{A} \rightarrow \text{J}/\psi \mathrm{A}),
\label{crossSecPro}
\end{equation}
where $\omega_{\gamma}$ represents the photon energy, and $\sigma(\gamma \mathrm{A} \rightarrow \text{J}/\psi \mathrm{A})$ is the photonuclear cross section for \jpsi production. The photon flux, $dN_{\gamma}(\omega_{\gamma})/d\omega_{\gamma}$, is derived from the Weizsäcker-Williams approximation~\cite{Krauss:1997vr}:
\begin{equation}
\begin{aligned}
\frac{d^3 N_\gamma\left(\omega_\gamma, \vec{x}_{\perp}\right)}{d \omega_\gamma d \vec{x}_{\perp}} & =\frac{4 Z^2 \alpha}{\omega_\gamma}\left|\int \frac{d^2 \vec{k}_{\gamma \perp}}{(2 \pi)^2} \vec{k}_{\gamma \perp} \frac{F_\gamma\left(\vec{k}_\gamma\right)}{\left|\vec{k}_\gamma\right|^2} e^{i \vec{x}_{\perp} \cdot \vec{k}_{\gamma \perp}}\right|^2, \\
\vec{k}_\gamma & =\left(\vec{k}_{\gamma \perp}, \frac{\omega_\gamma}{\gamma_c}\right), \quad \omega_\gamma=\frac{1}{2} M_{\mathrm{J} / \psi} e^{ \pm y},
\end{aligned}
\label{photoFulx}
\end{equation}
where $Z$ is the nuclear charge, $\alpha$ is the fine-structure constant, and $F_{\gamma}(\vec{k}_{\gamma})$ is the nuclear form factor obtained from the Fourier transform of the nuclear charge density. The variables $M_{\text{J}/\psi}$ and $y$ denote the mass and rapidity of the \jpsi, respectively. The term $\vec{x}_{\bot}$ corresponds to the photon's position transverse to the beam axis, and $\vec{k}_{\gamma \bot}$ is its momentum in the transverse plane. The Lorentz factor of the photon-emitting nucleus is denoted by $\gamma_{c}$.

The nuclear charge density $\rho_A(r)$, modeled by the Woods-Saxon distribution, is given by:
\begin{equation}
\rho_{A}(r) = \frac{\rho^{0}}{1 + \exp[(r - R_{\rm{WS}})/d]},
\end{equation}
where $\rho^0$ represents the normalization factor, while $R_{\rm{WS}}$ and $d$ are the radius and skin depth parameters, respectively. The specific values used in our calculation are provided in Table~\ref{tab:summary_table}.
\begin{table}[h]
  \centering
  \begin{tabular}{c c c}
    \hline\hline
    Nucleus  & $d \,(\mathrm{fm})$ & $R_{\mathrm{WS}} \,(\mathrm{fm})$\\ \hline
    Au   & 0.535  & 6.38   \\ 
    Pb   & 0.546  & 6.62  \\ \hline\hline
  \end{tabular}
  \caption{A summary table of the parameters used for the Woods-Saxon distribution of Au and Pb, respectively. The data are taken from the electron scattering experiments~\cite{barrett1977nuclear}.}
  \label{tab:summary_table}
\end{table}


The photonuclear cross section $\sigma(\gamma \mathrm{A} \rightarrow \text{J}/\psi \mathrm{A})$ is computed through the convolution of the differential cross section at $t=0$, where $-t$ represents the squared four-momentum transfer quantifying the momentum exchanged between the photon and the target (the nucleus or proton), and the nuclear form factor for the Pomeron~\cite{Zha:2017jch,PhysRevC.93.044912,PhysRevC.60.014903}:
\begin{equation}
\begin{aligned}
\sigma(\gamma \mathrm{A} \rightarrow \text{J}/\psi \mathrm{A}) = \\[0.3cm]
\frac{d\sigma(\gamma \mathrm{A} \rightarrow \text{J}/{\psi} \mathrm{A})}{dt} \bigg|_{t=0} \times \int |F_{P}(\vec{k}_{P})|^2 d^2 \vec{k}_{P\bot},
\end{aligned}
\label{crossSecGammaA}
\end{equation}
where $\vec{k}_P=\left(\vec{k}_{P \perp}, \frac{\omega_P}{\gamma_c}\right)$ is the Pomeron momentum, and $\omega_P=\frac{M_{J / \psi}^2}{4\omega_{\gamma}}$ is the energy associated with the Pomeron. The form factor $F_{P}(\vec{k}_{P})$ is extracted from the nuclear density, which is approximated by the nuclear charge density.
The differential cross section at $ t = 0 $ for the process $ \gamma \mathrm{A} \rightarrow \text{J}/\psi \mathrm{A} $ is described by the impulse approximation (IA)~\cite{Kryshen:2023bxy}:
\begin{equation}
\frac{d\sigma(\gamma \mathrm{A} \rightarrow \text{J}/\psi \mathrm{A})}{dt} \bigg|_{t=0} = \frac{d\sigma(\gamma \mathrm{p} \rightarrow \text{J}/\psi \mathrm{p})}{dt} \bigg|_{t=0} \mathrm A^2.
\end{equation}
The cross section for $ \gamma \mathrm{p} \rightarrow \text{J}/\psi \mathrm{p} $ is obtained from experimental measurements~\cite{Cao:2018hvy}. The following parametrization is used in our calculations~\cite{Cao:2018hvy}:
\begin{equation}
\begin{aligned}
\frac{d\sigma(\gamma \mathrm{p} \rightarrow \text{J}/\psi \mathrm{p})}{dt} \bigg|_{t=0} = \\[0.1cm]
\mathrm{C}_0\left(1-\frac{\left(m_{\mathrm{p}}+M_{\mathrm{J /} \psi}\right)^2}{\mathrm{W}_{\gamma \mathrm{N}}^2}\right)^{1.5}\left(\frac{\mathrm{W}_{\gamma \mathrm{N}}^2}{100^2 \mathrm{GeV}^2}\right)^\delta,
\end{aligned}
\end{equation}
where $ C_0 = 80.2 \pm 0.9 \, \mathrm{nb} / \mathrm{GeV}^2 $, $ \delta = 0.40 \pm 0.01 $, and $ m_{\mathrm{p}} $ and $ M_{\mathrm{J /} \psi} $ are the masses of the proton and $ \mathrm{J /} \psi $, respectively. This formula effectively fits data from RHIC and LHC experiments for the $ \gamma \mathrm{p} \rightarrow \mathrm{J /} \psi \mathrm{p} $ process.

The $ t $-dependence of the coherent photonuclear cross section, $ \phi_{\mathrm A}(t) $, is determined by the squared nuclear form factor:
$\phi_{\mathrm A}(t)=\frac{d\sigma(\gamma \mathrm{p} \rightarrow \text{J}/\psi \mathrm{p})}{dt} \bigg|_{t=0}\left|{\mathrm A} F_{\mathrm A}(t)\right|^2$. The $t$-integrated cross section can be expressed as~\cite{Klein:2016yzr}:
\begin{equation}
\Phi_{\mathrm A}\left(t_{\min }\right)=\int_{t_{\min }}^{\infty} d t\phi_{\mathrm A}(t),
\end{equation}
where $ t_{\text{min}} = - p_{z, \text{min}}^2 = -\left( \frac{M_{\mathrm{J /} \psi}^2}{4 \omega \gamma_L} \right)^2 $ is related to the minimum longitudinal momentum transfer $ p_{z, \text{min}} $, which governs the coherence length and becomes important at low energies. Here, $ \omega $ represents the photon energy, and $ \gamma_L $ is the Lorentz factor of the nucleus.

To take into account the nuclear effects for coherent \jpsi production in UPCs, we also incorporate gluon nPDF distributions. In this work, the EPPS21~\cite{Eskola:2021nhw} and nCTEQ15~\cite{Kovarik:2015cma} nPDF sets from LHAPDF6~\cite{Buckley:2014ana} are considered. The refined cross section calculation is given by:
\begin{equation}
\begin{aligned}
\sigma_{\text{AA} \rightarrow \text{AA} \text{J}/\psi}^{'}(y) = \\[0.3cm]
\sigma_{\text{AA} \rightarrow \text{AA} \text{J}/\psi}(x_1) \mathrm{R_{g}^{A}}(x_1)^2 + \sigma_{\text{AA} \rightarrow \text{AA} \text{J}/\psi}(x_2) \mathrm{R_{g}^{A}}(x_2)^2,
\end{aligned}
\label{refinedCrossSec}
\end{equation}
where $\mathrm{R_{g}^{A}}(x_{1,2})$ represent the nuclear gluon modification factors, and $x_{1,2}$ correspond to the two Bjorken $x$ values, defined as:
\begin{equation}
x = \frac{M_{\jpsi} e^{\pm y}}{\snn}.
\label{Bjox}
\end{equation}


\subsection{Bayesian reweighting}
To prepare for reweighting, it is necessary to convert the nPDF sets derived from Hessian error PDFs into a set of PDF replicas that represent the underlying probability distribution. These PDF replicas are defined using the formula:
\begin{equation}
f_k = f_0 + \sum_{i=1}^N \frac{f_i^{(+)} - f_i^{(-)}}{2} R_{k i},
\label{PDFReplica}
\end{equation}
where $f_{0}$ denotes the central PDF from the best fit, $f_{i}^{(+)}$ and $f_{i}^{(-)}$ are the PDFs corresponding to the positive and negative deviations along the eigenvector $i$, and $N$ represents the number of eigenvectors used to define the Hessian error PDFs. The term $R_{k i}$ is a random number drawn from a Normal distribution, unique for each replica $k$ and each eigen-direction $i$. This method allows for the generation of a diverse set of PDF replicas, which are crucial to accurately assess the impact of new data on the nPDFs.

After generating the PDF replicas, we can compute the average and variance of any observable ($\langle\mathcal{O}\rangle$ and $\delta\langle\mathcal{O}\rangle$) dependent on the PDFs by treating them as moments of the associated probability distribution:
\begin{equation}
\begin{aligned}
\langle \mathcal{O} \rangle & = \frac{1}{N_{\text{rep}}} \sum_{k=1}^{N_{\text{rep}}} \mathcal{O}(f_k), \\
\delta \langle \mathcal{O} \rangle & = \sqrt{\frac{1}{N_{\text{rep}}} \sum_{k=1}^{N_{\text{rep}}} \left( \mathcal{O}(f_k) - \langle \mathcal{O} \rangle \right)^2},
\end{aligned}
\end{equation}
where $\mathcal{O}(f_k)$ is the theoretical prediction from replica $k$.

Specifically, we can apply this approach to the PDFs themselves. The average of the replicas, $\langle f \rangle$, should yield our central PDF $f_0$, while the corresponding variance, defined as $\Delta f = \frac{1}{2} \sqrt{\sum_{i=1}^N \left( f_i^{(+)} - f_i^{(-)} \right)^2}$, would give us the Hessian error bands $\Delta \langle f \rangle$. The accuracy with which we can reproduce the central Hessian PDFs and their associated uncertainties is contingent upon how well we capture the underlying probability distribution, which in turn depends on the number of replicas, $N_{\text{rep}}$, utilized in the analysis. In this work, we adopt $N_{\text{rep}} = 10^4$, as demonstrated in~\cite{Kusina:2016fxy}, which facilitates a robust reproduction of both the central and error PDFs.

Having established the replicas, we can now apply the reweighting technique to assess the influence of a given data set on our current PDFs. This method is grounded in Bayes' theorem, which states that the posterior distribution—representing the probability of a hypothesis, in this case a new probability distribution obtained by incorporating the new data set into our analysis—is proportional to the product of the prior probability (the original PDFs without the new data) and a suitable likelihood function. This framework allows us to assign a weight to each of the previously generated replicas according to Equation (\ref{PDFReplica}).
The weight is defined as~\cite{zhang2021modern}:
\begin{equation}
w_k = \frac{\left(\chi_k^2\right)^{\frac{1}{2}(n-1)} e^{-\chi_k^2 / 2}}{\frac{1}{N_{\text{rep}}} \sum_{k=1}^{N_{\text{rep}}} \left(\chi_k^2\right)^{\frac{1}{2}(n-1)} e^{-\chi_k^2 / 2}},
\end{equation}
where $n$ denotes the number of measurements and $\chi^2_k$ represents the chi-squared statistic for the data sets considered in the reweighting procedure for the specific replica $k$. This chi-squared statistic can be expressed as~\cite{zhang2021modern}:
\begin{equation}
\chi_k^2(Y, f_k) = \sum_{i,j=1}^n \left(Y_i - Y_i[f_k]\right) \operatorname{cov}_{ij}^{-1} \left(Y_j - Y_j[f_k]\right),
\end{equation}
where $Y$ denotes the measurements, and $Y[f_k]$ is the theoretical prediction calculated with PDF sets derived from replica $k$.

With this framework in place, we can compute the weights necessary for the reweighting procedure. Subsequently, we can evaluate the expectation value ($\langle \mathcal{O} \rangle$) and variance ($\delta \langle \mathcal{O} \rangle_{\text{new}}$) of any observable dependent on the PDFs using weighted sums:\
\begin{equation}
\begin{aligned}
\langle \mathcal{O} \rangle_{\text{new}} & = \frac{1}{N_{\text{rep}}} \sum_{k=1}^{N_{\text{rep}}} w_k \mathcal{O}(f_k), \\
\delta \langle \mathcal{O} \rangle_{\text{new}} & = \sqrt{\frac{1}{N_{\text{rep}}} \sum_{k=1}^{N_{\text{rep}}} w_k \left( \mathcal{O}(f_k) - \langle \mathcal{O} \rangle_{\text{new}} \right)^2}.
\end{aligned}
\end{equation}

\begin{figure*}[ht!]
  \centering
  \includegraphics[width=.95\linewidth]{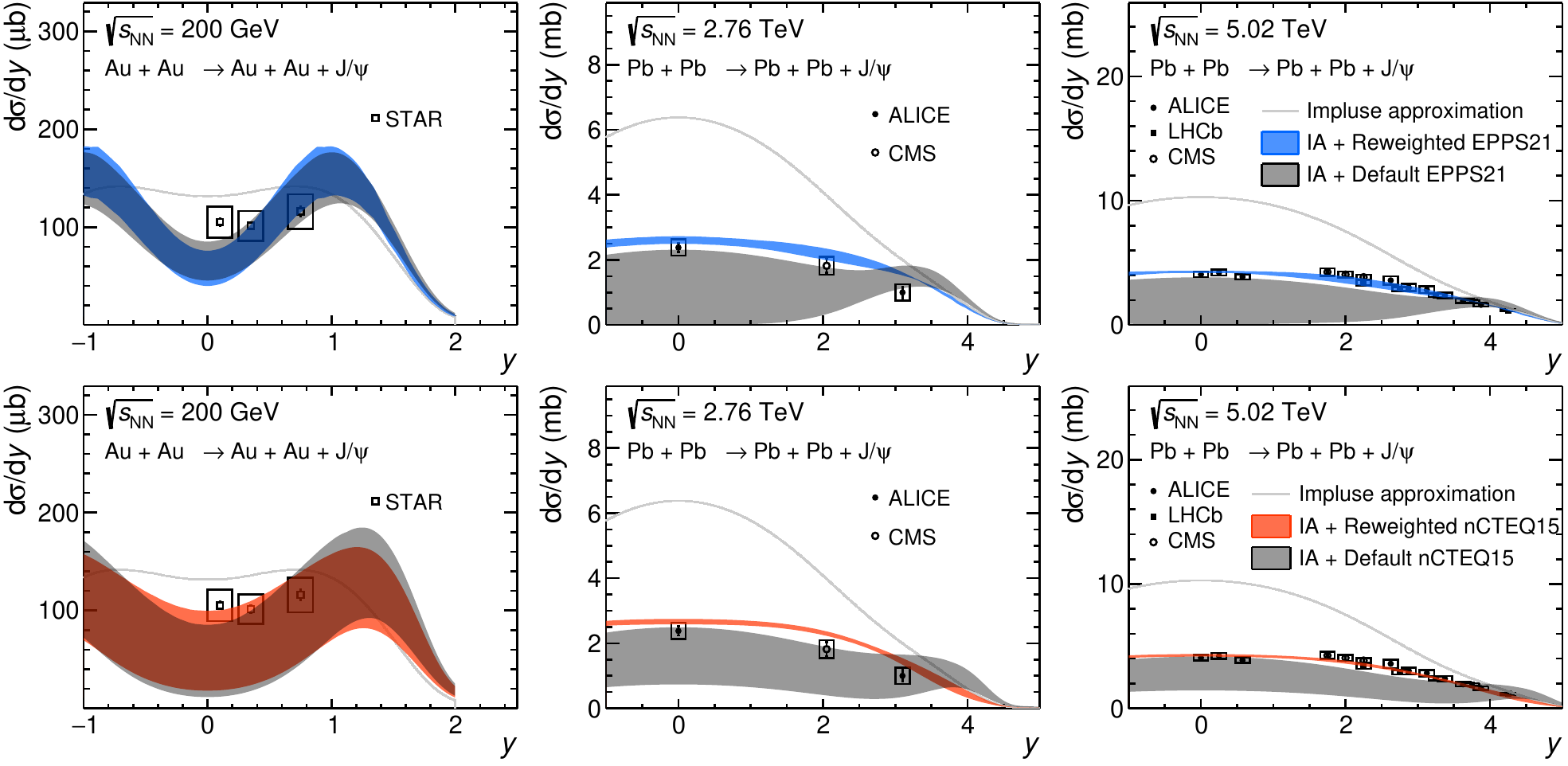}
  \caption{
  The $y$-differential cross section measurements of coherent photon-induced \jpsi production in Au--Au UPCs at \twoHnn~\cite{STAR:2023vvb,STAR:2023nos} and in Pb--Pb UPCs at \twosevensixnn~\cite{ALICE:2012yye,CMS:2016itn} and at \fivenn~\cite{ALICE:2021gpt,ALICE:2019tqa,LHCb:2021bfl,CMS:2023snh}. From left to right, each column showcases one of the three collision energies under consideration. Within each panel, we compare the experimental cross section with the predicted cross section from the impulse approximation (neglecting gluon nPDF effects) alongside the predictions that incorporate nPDF effects after the Bayesian reweighting. The top and bottom rows represent the EPPS21~\cite{Eskola:2021nhw} and nCTEQ15~\cite{Kovarik:2015cma} nPDF sets, respectively. The grey bands show the default nPDF versions, while the colored bands illustrate the reweighted nPDF sets.
}
  \label{crossSecVsRap}
\end{figure*}

\section{Results and discussion}
\label{results}


Building on the described methodology, we analyze rapidity-differential cross section measurements from coherent \jpsi production to better constrain the gluon nPDFs within Pb and Au nuclei. The experimental uncertainties are treated as follows: systematic uncertainties within each individual measurement are considered fully correlated across rapidity bins, reflecting their common experimental origin, while statistical uncertainties remain uncorrelated. This treatment ensures proper propagation of experimental uncertainties in our Bayesian reweighting procedure. 

Figure~\ref{crossSecVsRap} shows the $y$-differential cross section measurements of coherent photon-induced \jpsi production without neutron tagging in Au--Au UPCs at \twoHnn, as reported by STAR~\cite{STAR:2023vvb,STAR:2023nos}, and in Pb--Pb UPCs at \twosevensixnn by ALICE~\cite{ALICE:2012yye} and CMS~\cite{CMS:2016itn}. Data from \fivenn Pb--Pb UPCs, measured by ALICE~\cite{ALICE:2021gpt,ALICE:2019tqa}, LHCb~\cite{LHCb:2021bfl}, and CMS~\cite{CMS:2023snh}, are also included. Each column in the figure corresponds to one of the three collision energies considered. The two evaluated nPDF sets, EPPS21~\cite{Eskola:2021nhw} and nCTEQ15~\cite{Kovarik:2015cma}, are presented by different colors.

In the IA, no nPDF effects are considered. Predictions from the IA model significantly overestimate the measurements at all three energies, particularly in Pb--Pb UPCs, from forward rapidity to midrapidity. At LHC energies, \jpsi production at midrapidity probes very small gluon Bjorken $x$, where strong nuclear shadowing is expected. At forward rapidity, larger $x$ values lead to a reduced discrepancy between measurements and IA predictions, reflecting weaker shadowing. A similar pattern of shadowing is evident at RHIC energy at midrapidity (left panel).

Following Equation (\ref{refinedCrossSec}), one can predict coherent \jpsi production considering the default nPDFs, which are shown in grey bands. Despite the significant uncertainties associated with the default nPDFs, these predictions are capable of describing the experimental data. To refine the gluon momentum distribution within the Pb nucleus, we performed a global Bayesian reweighting using the combined \twosevensixnn~\cite{ALICE:2012yye,CMS:2016itn} and \fivenn~\cite{ALICE:2021gpt,ALICE:2019tqa,LHCb:2021bfl,CMS:2023snh} datasets, incorporating all available measurements. Similarly, for the Au nucleus, a Bayesian reweighting was performed using the \twoHnn datasets~\cite{STAR:2023vvb,STAR:2023nos}. The reweighted predictions, shown as colored bands in Figure~\ref{crossSecVsRap}, correspond to the EPPS21 and nCTEQ15 nPDF sets. These bands represent the cross sections after reweighting.


\begin{figure*}[ht!]
  \centering
  \includegraphics[width=.95\linewidth]{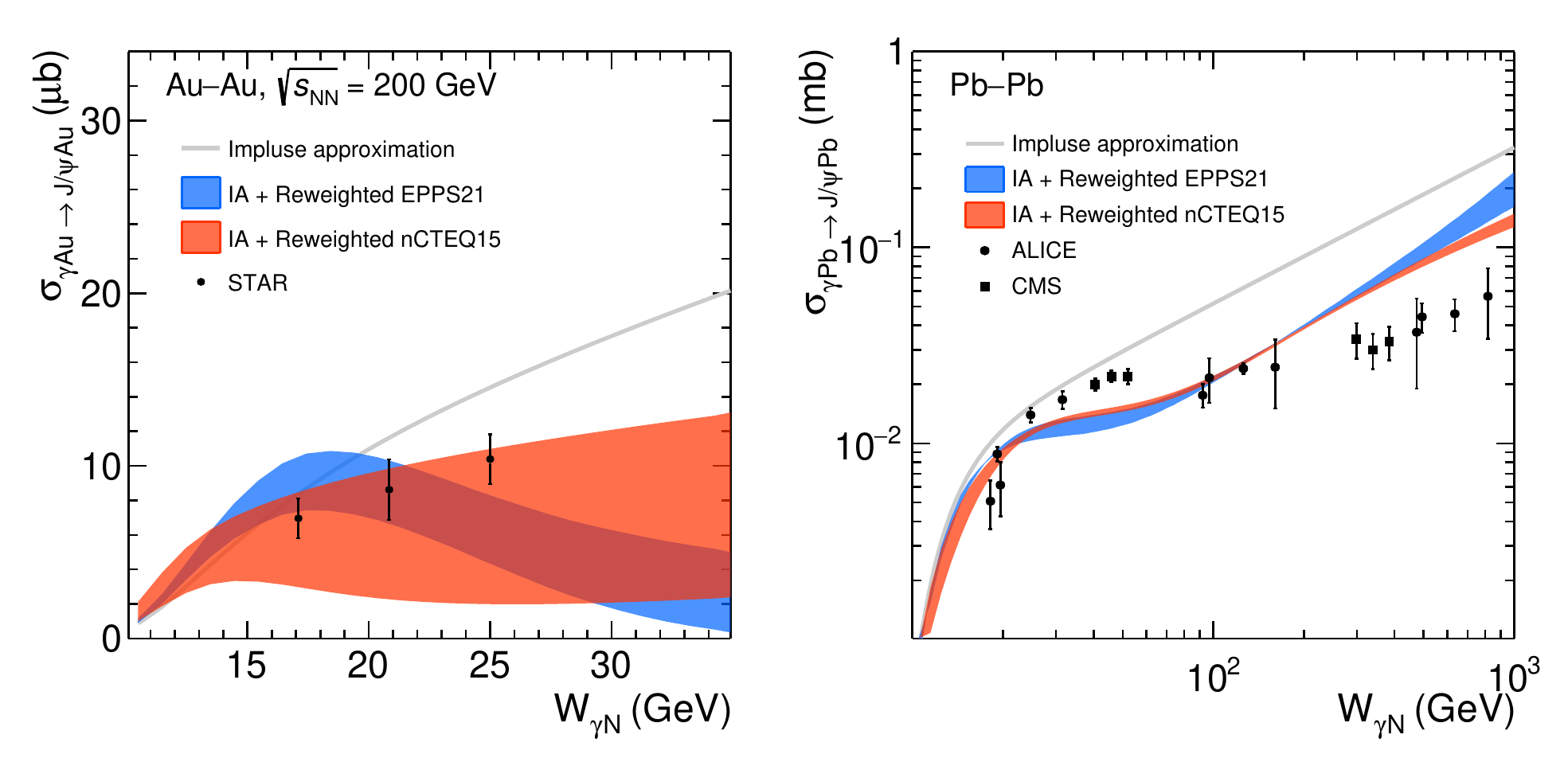}
  \caption{
  The integrated coherent \jpsi photoproduction cross section as a function of $\mathrm{W}_{\gamma \mathrm{N}}$ for Au--Au at \twoHnn (left) and Pb--Pb at \twosevensixnn and \fivenn (right) UPCs. The $\mathrm{W}_{\gamma \mathrm{N}}$ values correspond to the center of each experiment's rapidity range. Measured results are represented by black points, with vertical bars indicating the total experimental uncertainties~\cite{STAR:2023vvb,ALICE:2023jgu,CMS:2023snh}. Predictions from the impulse approximation are displayed as grey lines, while the reweighted nPDFs are illustrated by colored bands.
}
  \label{crossSecVsWgammaN}
\end{figure*}

\begin{figure*}[ht!]
  \centering
  \includegraphics[width=.95\linewidth]{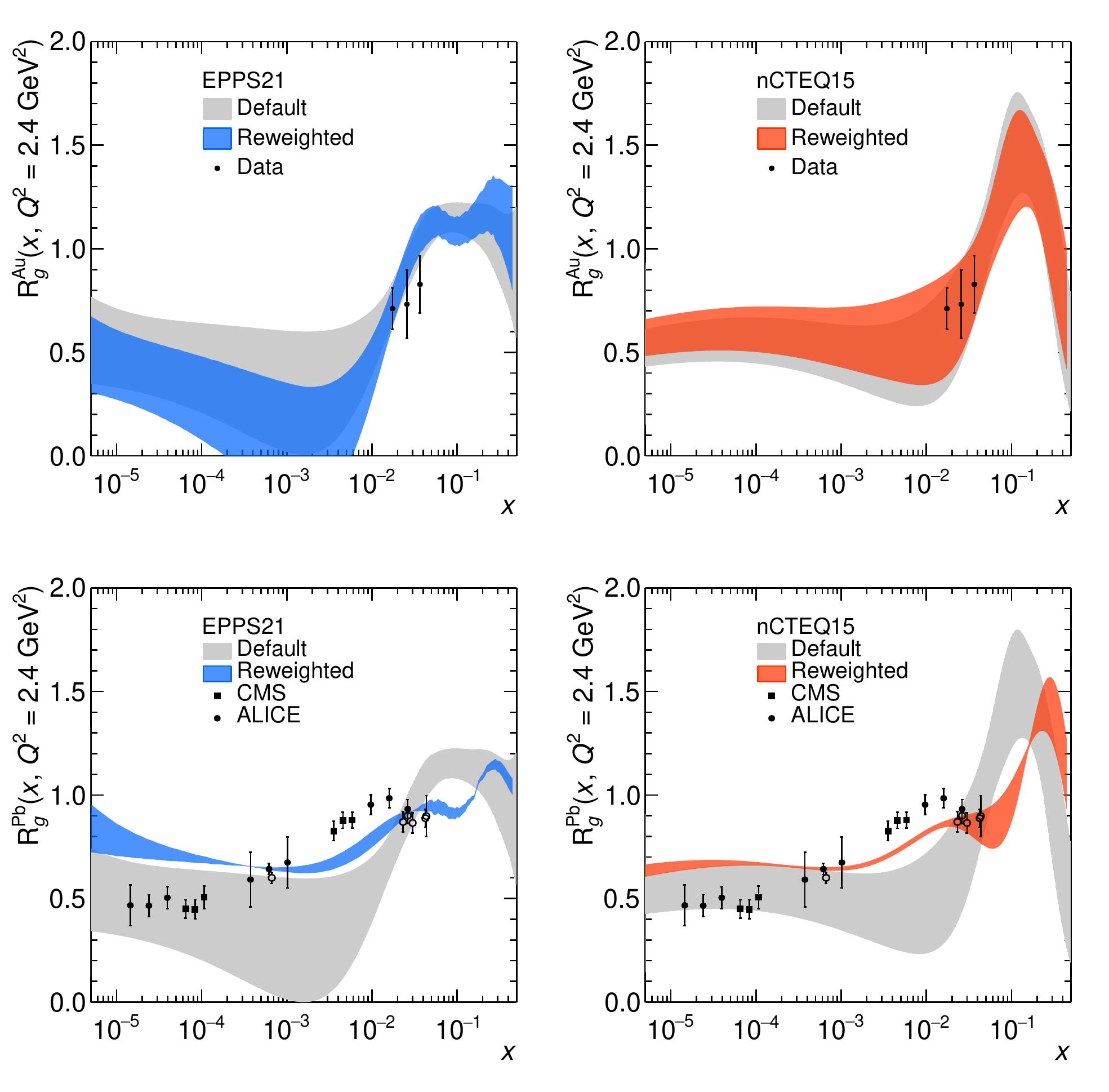}
  \caption{
  The nuclear gluon suppression factors $\mathrm{R_{g}^{Au}}$ (top) and $\mathrm{R_{g}^{Pb}}$ (bottom) as a function of Bjorken $x$, extracted from coherent \jpsi photoproduction in heavy-ion UPCs. The panels, from left to right, correspond to the EPPS21 and nCTEQ15 nPDF sets, respectively. Solid points indicate the measured results at the RHIC~\cite{STAR:2023nos} and the LHC~\cite{ALICE:2023jgu,CMS:2023snh}, with vertical lines extending across each marker representing the total uncertainties, calculated as the quadrature sum of statistical and systematic uncertainties. The open circles are the approximated results, where the two-way ambiguity effect is expected to be negligible, from the ALICE~\cite{ALICE:2019tqa,ALICE:2021gpt}, LHCb~\cite{LHCb:2022ahs} and E691~\cite{PhysRevLett.57.3003,Guzey:2020ntc} experiments. The nPDF distributions before reweighting are shown as grey bands, while the updated nPDFs are displayed in blue (EPPS21) and orange (nCTEQ15).
}
  \label{suppressionFactor}
\end{figure*}

Overall, the reweighted predictions--both EPPS21 (blue) and nCTEQ15 (orange)--closely match the measurements across all rapidities for the three collision energies. Compared to Au, the extensive coverage and precision of the Pb--Pb UPC data enable these refined predictions to achieve significantly reduced uncertainties.

As illustrated in Figure~\ref{crossSecVsWgammaN}, to gain deeper insights into gluon nuclear distribution functions, we use the reweighted distributions to predict the total \jpsi coherent photoproduction cross section as a function of $\mathrm{W}_{\gamma \mathrm{N}}$ for both Au--Au and Pb--Pb UPCs. These predictions are compared with RHIC and LHC experimental measurements obtained using neutron tagging techniques, as mentioned in the introduction and further detailed in the following.
As outlined in Equation (\ref{photoFulx}), due to the symmetry inherent in A--A collision systems, a \jpsi meson measured at rapidity $y$ can be associated with two distinct photon energies. For instance, given $\omega_\gamma = \frac{1}{2} M_{\mathrm{J}/\psi} e^{-y}$, the Bjorken $x$ of the parton and $\mathrm{W}{\gamma \mathrm{N}}$ can be calculated as:
$x=\left(M{\mathrm{J}/\psi} / \sqrt{s_{\mathrm{NN}}}\right) e^{+y}$ and $\mathrm{W}{\gamma \mathrm{N}}^{\mathrm{A}} = \sqrt{s{\mathrm{NN}}} M_{\mathrm{J}/\psi} e^{-y}$.
By reversing the sign of $y$, one can determine the values of $x$ and $\mathrm{W}{\gamma \mathrm{N}}$ corresponding to the higher photon energy. This two-fold ambiguity in $x$ or $\mathrm{W}{\gamma \mathrm{N}}$ has been addressed in recent publications from ALICE~\cite{ALICE:2023jgu}, CMS~\cite{CMS:2023snh}, and STAR~\cite{STAR:2023vvb,STAR:2023nos} by employing neutron tagging techniques. These techniques effectively constrain the impact parameter range in ultra-peripheral collisions, creating distinct photon fluxes as a function of energy, and thereby resolve the two-component ambiguity in \jpsi photoproduction in heavy-ion collisions.
However, this approach relies heavily on theoretical models to calculate the neutron emission probability. Currently, the reliability and uncertainty of theoretical calculations for Coulomb dissociation remain insufficiently quantified. Moreover, these calculations typically assume that the excitation of the two nuclei is independent and that the dissociation process is uncorrelated with the production process. These assumptions of independence have not been rigorously validated through theoretical calculations or verified experimentally, introducing potential biases into the results.

For the Au--Au collision system (shown on the left), due to the relatively large experimental uncertainties, the measured data (using neutron tagging) are consistent with the IA approach, as well as with the predictions from both the default and reweighted PDF sets. However, for the Pb--Pb collision system (shown on the right), both the experimental data and the predictions from the reweighted PDF sets are systematically lower than those from the IA approach, indicating a significant shadowing effect for Pb nuclei. The experimental data and reweighted PDF predictions show good agreement in the intermediate Bjorken-$x$ range. This is expected, as the precision in this range is primarily determined by measurements at mid-rapidity ($y=0$) for coherent \jpsi production, where the two-component ambiguity is absent.

In contrast, significant discrepancies emerge in the low-$x$ and high-$x$ regions, which correspond to forward rapidities. In these regions, the predictions from the reweighted PDF sets differ substantially from the experimental measurements. This inconsistency arises from the fact that the reweighted PDF method and neutron tagging experiments yield conflicting estimates of the low-$x$ and high-$x$ contributions at the same rapidity. Neutron tagging tends to assign a higher contribution to low-$x$.One possible explanation for these discrepancies is that the parameterizations of EPPS21 and nCTEQ15, particularly EPPS21, which provides a high degree of flexibility, might still not be sufficient to accurately describe the data in these extreme regions. Another explanation is that the neutron tagging method, which relies on Coulomb dissociation calculations, requires more robust theoretical modeling. These models should include considerations of correlations between nuclear excitations, as well as the higher-order effects induced by the photoproduction process itself. And cross sections extracted using neutron tagging techniques fail to adequately describe the coherent \jpsi production measurements at forward rapidities in \twosevensixnn Pb--Pb collisions. A more direct and robust method to resolve these discrepancies would involve extending precise measurements of coherent \jpsi production at mid-rapidity (where the two-component ambiguity is absent) to cover higher and lower energy ranges, corresponding to low-$x$ and high-$x$ regions, respectively.
These inconsistencies also call for caution when drawing conclusions from experimental results derived using neutron tagging, such as those related to the onset of the Color Glass Condensate (CGC) or the black-disk limit.

Figure~\ref{suppressionFactor} illustrates the reweighted nuclear gluon modification factor, $\mathrm{R_{g}^{Au, Pb}}(x,, Q^2=2.4,\mathrm{GeV^2})$, plotted as a function of $x$, where $x = M_{\jpsi}^{2}/W_{\gamma \mathrm{N}}$ and $Q = M_{\jpsi}/2$. This factor is compared with the default predictions from nPDFs~\cite{Eskola:2021nhw, Kovarik:2015cma} and the experimental data obtained using neutron tagging technology~\cite{STAR:2023vvb, STAR:2023nos, ALICE:2023jgu, CMS:2023snh}. Experimentally, $\mathrm{R_{g}^{A}}$ is defined as $\sqrt{\sigma^{\mathrm{Meas}} / \sigma^{\mathrm{IA}}}$, where $\sigma^{\mathrm{Meas}}$ represents the measured cross section and $\sigma^{\mathrm{IA}}$ corresponds to the impulse approximation prediction.
The experimental data points from ALICE~\cite{ALICE:2023jgu} and CMS~\cite{CMS:2023snh} for Pb--Pb UPCs at \fivenn are represented by circular and square markers, respectively. Open circles denote approximations provided by CMS~\cite{CMS:2023snh}, which combine measurements from ALICE~\cite{ALICE:2021gpt}, LHCb~\cite{LHCb:2022ahs}, and E691~\cite{PhysRevLett.57.3003, Guzey:2020ntc}, where the impact of the two-component ambiguity is expected to be negligible.


For the Au nucleus, due to the limited availability of \jpsi coherent production data, the reweighted nPDFs exhibit only modest updates and adequately describe the experimental measurements. In contrast, for the Pb nucleus, the abundance of measurements across a wide rapidity ($y$) range provides broader coverage of the gluon suppression factor over a wide range of $x$. To facilitate the discussion, the $x$ range is divided into four regions: low-$x$ ($x < 2 \times 10^{-4}$), intermediate-$x$ ($2 \times 10^{-4} < x < 3 \times 10^{-3}$), relatively high-$x$ ($3 \times 10^{-3} < x < 2 \times 10^{-2}$), and high-$x$ ($x > 2 \times 10^{-2}$).
In the low-$x$ region, a suppression factor of approximately 0.6 is observed from both reweighted nPDF sets. As $x$ increases, the suppression factor rises to around 0.8 in the relatively high-$x$ and high-$x$ regions. This trend is consistent with the well-known shadowing effects in nuclear matter, while also achieving a significant reduction in gluon density uncertainties compared to the initial nPDF predictions.

When comparing the reweighted nPDFs with measurements obtained using neutron tagging, the conclusions align with observations in Figure~\ref{crossSecVsWgammaN}. Both the EPPS21 and nCTEQ15 models provide reasonable descriptions in the intermediate-$x$ and high-$x$ regions. However, discrepancies arise in the low-$x$ and relatively high-$x$ regions, where the reweighted nPDFs tend to overshoot and undershoot the experimental data, respectively. These discrepancies have been discussed in detail in previous sections.
One important point to emphasize is that the modification factors extracted using neutron tagging tend to exhibit a different trend compared to measurements from E691, where the two-component ambiguity is absent. This discrepancy underscores the potential limitations of the neutron tagging technique in accurately resolving contributions across different $x$ ranges.

\section{Summary}
\label{summary}
In this study, Bayesian reweighting is applied to the EPPS21 and nCTEQ15 nuclear parton distribution function (nPDF) sets by incorporating coherent \jpsi photoproduction measurements from RHIC and LHC. For the Au nucleus, the limited availability of \jpsi coherent production data results in only modest updates to the reweighted nPDFs, which nonetheless provide an adequate description of the experimental measurements. In contrast, the Pb nucleus benefits from an extensive dataset spanning a broad rapidity range, enabling improved coverage of gluon suppression factors across a wide $x$ range and yielding more precise constraints on the nuclear parton distributions.

The Bayesian-reweighted gluon modification factors, $\mathrm{R_g^{A}}(x, Q^2 = 2.4\ \mathrm{GeV}^2)$, reveal pronounced nuclear shadowing effects in Pb nuclei, with $\mathrm{R_g^{\mathrm{Pb}}} \approx 0.60$ at $x = 10^{-4}$. Compared to the initial nPDF predictions, the reweighted analysis significantly reduces uncertainties in the gluon density across the critical Bjorken-$x$ range of $10^{-5} < x < 10^{-3}$.
While the reweighted nPDFs show good agreement with experimental data obtained via neutron tagging in the intermediate-$x$ and high-$x$ regions, notable discrepancies are observed in the low-$x$ and relatively high-$x$ regions for Pb nuclei. These discrepancies point to potential limitations of the neutron tagging technique, particularly in resolving the two-component ambiguity in forward rapidity regions, and emphasize the need for further theoretical and experimental efforts to refine these methodologies.
Overall, this study highlights the effectiveness of Bayesian reweighting in reducing uncertainties and provides valuable insights into the gluon distribution functions of heavy nuclei, especially in the shadowing-dominated regime.




\textit{Acknowledgment}
{The authors would like to thank Prof. Shuai Yang, Prof. Zaochen Ye, Prof. Zhangbu Xu, Prof. Hongxi Xin, Prof. Jian Zhou and many members of the STAR, ATLAS, ALICE, CMS, and LHCb Collaborations. This work is supported in part by the National Key Research and Development Program of China under Contract No. 2022YFA1604900 and the National Natural Science Foundation of China (NSFC) under Contract No. 12175223 and 12005220. W. Zha is supported by Anhui Provincial Natural Science Foundation No. 2208085J23 and Youth Innovation Promotion Association of Chinese Academy of Science.}

\bibliography{bibliography}

\end{document}